\documentclass[12pt, a4paper]{article}

\renewcommand{\thefootnote}{\fnsymbol{footnote}}
\usepackage{amssymb}
\usepackage{amsmath}
\usepackage{amsfonts}
\usepackage{graphicx}
\usepackage{mathrsfs}
\usepackage{comment}
\def\->{\rightarrow}
\def\<-{\leftarrow}

\newcommand{\Slash}[1]{{\ooalign{\hfil#1\hfil\crcr\raise.167ex\hbox{/}}}}

\newcommand{\beq}{\begin{equation}}  \newcommand{\eeq}{\end{equation}}
\newcommand{\bef}{\begin{figure}}  \newcommand{\eef}{\end{figure}}
\newcommand{\bec}{\begin{center}}  \newcommand{\eec}{\end{center}}
\newcommand{\non}{\nonumber}  \newcommand{\eqn}[1]{\beq {#1}\eeq}
\newcommand{\laq}[1]{\label{eq:#1}}  

\newcommand{\Eq}[1]{Eq.(\ref{eq:#1})}
\newcommand{\Eqs}[1]{Eqs.(\ref{eq:#1})}
\newcommand{\eq}[1]{(\ref{eq:#1})}
\newcommand{\Sec}[1]{Sec.\ref{chap:#1}}
\newcommand{\ab}[1]{\left|{#1}\right|}
\newcommand{\vev}[1]{ \left\langle {#1} \right\rangle }

\newcommand{\lac}[1]{\label{chap:#1}}
\newcommand{\SU}[1]{{\rm SU{#1} } }
\newcommand{\SO}[1]{{\rm SO{#1}} }
\newcommand{\bed}{\begin{description} \item}
\newcommand{\eed}{\end{description}}

\def\({\left(}
\def\){\right)}
\def\dt{{d \o dt}}

\def\O{\mathcal{O}}
\def\U{\mathop{\rm U}}

\def\tr{\mathop{\rm tr}}

\newcommand{\OR}{~{\rm or}~}
\newcommand{\AND}{~{\rm and}~}

\newcommand{\GEV}{ {\rm GeV} }
\newcommand{\TEV}{ {\rm TeV} }

\def\o{\over}
\def\a{\alpha}
\def\b{\beta}

\def\d{\delta}
\def\e{\epsilon}
\def\f{\phi}
\def\g{\gamma}
\def\h{\theta}

\def\l{\lambda}
\def\m{\mu}
\def\n{\nu}
\def\p{\psi}

\def\s{\sigma}

\def\w{\omega}
\def\x{\xi}

\def\ol{\overline}
\def\tl{\tilde}
\def\*{\dagger}

\begin{document}
\begin{titlepage}
\begin{center}

\hfill TU-1029\\

\vspace{1.0cm}

{\Large\bf 
Fixed Point and Anomaly Mediation in \\
Partially $N=2$ Supersymmetric Standard Models}
\vspace{1.0cm}

{ {\bf  Wen Yin 
\footnote[0]{ email:  wyin@ihep.ac.cn}}}

 \vspace{1.0cm}
{\it Department of Physics, Tohoku University, Sendai 980-8578, Japan\\} 
{\it Institute of High Energy Physics, Chinese Academy of Science, \\
Beijing 100049, China}
\vspace{1.0cm}


\begin{abstract}
Motivated by the simple toroidal compactification of extra-dimensional SUSY theories, 
we investigate on a partially $N=2$ supersymmetric (SUSY) extension of the standard model which has an $N=2$ SUSY sector and an $N=1$ SUSY sector. 
We point out that below the scale of the partially breaking of $N=2$ to $N=1$, the ratio of Yukawa to gauge couplings embedded in the original $N=2$ gauge interaction in the $N=2$ sector becomes greater due to a fixed point. 
Since at the partially breaking scale the sfermion masses in the $N=2$ sector is suppressed due to the $N=2$ non-renormalization theorem, the anomaly mediation effect becomes important.
If dominant, the anomaly induced masses for the sfermions in the $N=2$ sector are almost UV-insensitive due to the fixed point. 
Interestingly, these masses are always positive, i.e. no tachyonic slepton problem. 
From an example model, we show interesting phenomena differing from the ordinary MSSM. In particular, the dark matter can be a sbino, i.e. the scalar component of the $N=2$ vector multiplet of $\U(1)_Y$.  To obtain the correct dark matter abundance, the mass of sbino, as well as the MSSM sparticles in the $N=2$ sector which have a typical mass pattern of anomaly mediation, is required to be small. 
Therefore, this scenario can be tested and confirmed in the LHC and may be further confirmed by the measurement of the $N=2$ Yukawa couplings in future colliders.
This model can explain dark matter, muon $g-2$ anomaly, gauge coupling unification and relaxes some ordinary problems within the MSSM. Also, it is compatible with thermal leptogenesis.
\end{abstract}




\end{center}
\end{titlepage}
\setcounter{footnote}{0}

\setcounter{footnote}{0}
\renewcommand{\thefootnote}{\arabic{footnote}}

\section{Introduction}

\lac{intro}
The supersymmetric (SUSY) extension of the Standard Model (SM) is a leading candidate of the new physics 
above TeVs, which can explain the discrepancy between the soft SUSY breaking and the 
fundamental scales. Theoretically, SUSY is needed to construct a self-consistent quantum gravity~\cite{Polchinski}.

However, from a theoretical point of view, 
there is no reason that the minimal SUSY extension of the SM (MSSM) is the physics next to the electroweak (EW) scale. Whereas, the SUSY SM may include more particle contents.
In particular, $N=1$ SUSY in higher dimensional space-time, such as the effective theory of superstring~\cite{Polchinski}, with a simple toroidal compactification is represented as $N=2$ SUSY~\cite{Grimm:1977xp} in the four-dimensional space-time~\cite{PN2UV2}\footnote{Lagrangian with $N>2$ SUSY can be represented by $N=2$ SUSY.}. 
Phenomenologically, $N=2$ SUSY should be partially broken down to $N=1$ and chirality appears at 
a scale, say $M_{\rm p}\simeq 2\times 10^{16}\GEV$.
If the partially breaking takes place only at a sector ($N=1$ sector), with the other 
sector ($N=2$ sector) sequestered, some of the $N=2$ SUSY partners ($N=2$ partners) of the SM in the $N=2$ sector may have masses around the soft SUSY mass scale which is supposed to be around TeVs.

This possibility gives several alternative experimental features to the ordinary MSSM case \cite{N2ph}.
In this paper, we investigate on the possibility that the $N=2$ vector and hyper- partners all remain until TeVs.
In particular, we will show that in this setup at a sufficiently low energy scale the gauge and Yukawa ($N=2$ Yukawa) couplings, originated from the $N=2$ gauge interaction for an $N=2$ hypermultiplet (hypermultiplet), are related by the fixed point of the renormalization group (RG). The ratio of the Yukawa to gauge couplings is raised by the RG effect towards this fixed point from the ratio at the $N=2$ SUSY limit. Namely, we predict typical Yukawa couplings for these hypermultiplets, which depend less on the partially breaking mechanism as well as the threshold corrections at $M_{\rm p}$.

Suppose that the soft breaking terms are generated above the partially breaking scale, the soft breaking mass squares of the hypermultiplets are vanishing due to the $N=2$ non-renormalization theorem~\cite{Shimizu:2015ara}.  
Then the soft masses for the sfermions in the $N=2$ sector below the partially breaking scale are generated via radiative corrections. One of the corrections is from the anomaly mediation effect~\cite{ANSBLS,ANSBM}. 
Since is a function of the particle's couplings and the gravitino mass, the anomaly induced masses in the $N=2$ sector are almost UV-insensitive due to the fixed point. Interestingly, due to the large ratio of the Yukawa and gauge couplings at the fixed point the anomaly induced mass squares for the hypermultiplets are positive. Namely, the tachyonic slepton problem\footnote{There are typically two possibilities to solve this problem: (a) the anomaly mediation effect is 
canceled or negligible~\cite{ANSBLS, Pomarol:1999ie, Shimizu:2015ara, Yin:2016shg}, and (b) the 
sfermions have large Yukawa interactions~\cite{Chacko:1999am}. The solution here belongs to the latter one.} in the $N=2$ sector is automatically solved near the fixed point.

In a concrete partially $N=2$ SSM, we confirm that the $N=2$ Yukawa couplings do converge to the fixed point at 2-loop level, and also the corresponding positive anomaly-induced slepton mass squares.
When the typical soft mass scale in the $N=1$ sector is around $\O(10)\TEV$, the radiative corrections from the top loops can be large enough to explain the Higgs boson mass~\cite{Okada:1990vk}. 
 Due to the light smuons and bino in the $N=2$ sector, we also find that the muon $g-2$ anomaly~\cite{muong2exp,muong2SM,muong2SM2} can be explained within its $1\s~$level error for the gravitino mass of $\O(100)\TEV$\footnote{For other explanations of the muon $g-2$ anomaly with heavy stops and light smuons, see \cite{Ibe:2012qu, Yin:2016shg}.}. 

The quite different perspective from the ordinary MSSM is dark matter. The ordinary neutralinos cannot be the candidate of dark matter because they are all heavier than the smuon due to the mass relation of anomaly mediation with additional multiplets.
The interesting candidates are the $N=2$ vector partners which are stabilized by a new $Z_2-$parity introduced to solve the tad-pole problem. 
The scalar component of the $N=2$ $\U(1)_Y$ vector multiplet, sbino, can explain the correct dark matter abundance when its mass is up to $700\GEV$ and the annihilation products,
some sleptons, are even lighter. 
Due to this mass constraints, if these light sleptons are long-lived enough, they can be fully tested in the LHC. 
The other $N=2$ sector sparticles of the MSSM has a typical mass patterns, which is related by anomaly mediation, with scale smaller than $\sim 2 \TEV$. 
This typical spectrum could be measured in the LHC.  Furthermore, depending on their mass range the $N=2$ partners could be tested in future colliders, such as SPPC, FCC, CLIC and a Muon Collider Higgs Factory 
which could further confirm our scenario \cite{CEPC}.

From the cosmological viewpoint, this scenario is favored because of the heavy gravitino which relaxes the gravitino problem~\cite{Kawasaki:2008qe}. 
In particular, the reheating temperature can be large enough while avoiding the overproduction of the produced dark matter. Thus our scenario is compatible with thermal leptogenesis~\cite{Fukugita:1986hr} and can produce the correct baryon asymmetry. CP and FCNC problems in the ordinary MSSM are also relaxed due to the heavy $N=1$ sector sfemions.

This paper is organized as follows.
In \Sec{1}, we introduce the partially $N=2$ SUSY models, and derive the fixed point for the $N=2$ Yukawa couplings.
In \Sec{2}, we explain the anomaly mediation effect on the fixed point and show the absence of the tachyonic slepton problem. Also, will be explained is the $N=2$ non-renormalization theorem. 
In \Sec{app}, we discuss the concrete example of partially SSMs and its several phenomenological and cosmological aspects.
\Sec{con} is devoted to discussion and conclusions.

\section{Fixed Point in Partially $N=2$ Supersymmetric Model}

\lac{1}

We will focus on the possibility that the $N=2$ SUSY is broken down to $N=1$ at a high energy scale $M_{\rm p}$ in the $N=1$ sector, while the $N=2$ sector remains approximately $N=2$ SUSY at this scale. 
This possibility is not peculiar because the following theoretical backgrounds exist.

$N=2$ to $N=1$ partially SUSY breaking can take place spontaneously as an $N=1$ SUSY gauge theory can be 
described by non-linear realized $N=2$ SUSY theories with chiral matters of any representations \cite{Bagger:1994vj}. In particular, $N=2$ SUSY non-linear abelian gauge models with electric and magnetic $N=2$ Fayet-Iliopoulos terms are proved to have such breaking~\cite{PN2UV1}. Thus, a sector, which does not directly couple to the $N=2$ SUSY gauge fields inducing the partially breaking, has the partially breaking only at the higher order and can be identified as the $N=2$ sector.  

The extra-dimensional theory with branes is also a candidate to realize this possibility. For example, $N=1$ SUSY on $R_{1,3} \times S_1$  
spacetime can be compactified into $d=4, ~N=2$ SUSY at the low energy \cite{PN2UV2}.  If ``our world" is localized on one of the four-dimensional branes perpendicular to the extra dimension, the brane fields have $d=4,~N=1$ SUSY while the compactified bulk fields have $d=4, ~N=2$ SUSY.
Ordinary, these $N=2$ partners of the MSSM particles are projected out by assuming an orbifold
parity, but this is not necessary. 
Therefore, at the compactification scale, we have two sectors an $N=2$ sector composed of the bulk fields and an $N=1$ sector composed of the brane fields.

In this section, we will introduce the partially $N=2$ SUSY model as the effective theory of the previous backgrounds (or others) and study the RG behavior for the dimensionless couplings in its $N=2$ sector.

\subsection{Introduction to Partially $N=2$ Supersymmetric Model}
\lac{def}

To simplify the discussion, let us consider an $\U(N_c)=\SU(N_c) \times U(1)$ Yang-Mills theory with partially $N=2$ SUSY defined by the following Lagrangian as a toy model.

\begin{equation}
\laq{toy}
\mathcal{L}=\int{d^4\theta \left( K_{N=2}+K_{N=1}\right)}+\int{d^2\theta W_{N=2}}+\int{d^2\theta W_{mass}}+\int{d^2\theta W_{H}}
\end{equation} \begin{equation}
\non
+{1\over 2}\left(\int{d^2\h \tr \left[\mathcal{W}^c \mathcal{W}^{c} \right]}+{1\over 2}\int{d^2\h {\mathcal{W} \mathcal{W}}}\right)+h.c.
\end{equation} \begin{equation}
\laq{KN2}
K_{N=2}=2\tr[{W^\dagger e^{-2g_cV_c} W e^{2g_cV_c}}]+\ab{\f_Y}^2+\sum_i^{N_F}{\left( L_i^\dagger e^{2 g_cV_c+2Y_i g V}L^i+\ol{L}_ie^{-2g_cV_c-2Y_i g V}\ol{L}^{i \dagger}\right)}, \end{equation} \begin{equation}
\laq{WN2}
W_{N=2}=-\sum_i^{N_F}{\left({\sqrt{2}}  \w_i g_c \ol{L}_i W L_i+{\sqrt{2}} \tl{Y}_i g  \ol{L}_i \f_Y L_i\right)}.
\end{equation} \begin{equation}
\laq{KN1}
K_{N=1}=\sum_i^{N_F}{S_i^\dagger e^{2g_cV_c+2Y_i g V} S^i}+\sum_a^{N_f}{H_a^\dagger e^{2g_cV_c+2Y^H_a g V} H^a}
\end{equation} \begin{equation}
\laq{mass}
 W_{mass}=-M_W \tr[W^2]-M \f_Y^2-\sum_{i}^{N_F}M_i \ol{L}_i S^i.
\end{equation}


Here, $\mathcal{W}^c$ ($\mathcal{W}$) and $V_c$ ($V$) are the field strength and the corresponding gauge 
multiplet of the SU($N_c$) (U(1)) with gauge coupling, $g_c$ ($g$), respectively; $W$ ($\f_Y$) is an $\SU(N_c)$ adjoint (singlet) chiral multiplet which stands for the $N=2$ vector partner (See \Sec{N2l}) of the $\SU(N_c)$ ($\U(1)$) gauge particle. 
$L_i$, $\ol{L}_j$, $S_k$ and $H_a$ are chiral multiplets with representations of $(r_i,Y_i)$,$(\ol{r}_j,-Y_j)$, $(r_k, Y_k)$ and $(r^{H
}_a, Y^H_a)$ under $(\SU(N_c), U(1))$, respectively, where $L_i$ stand for the MSSM matter multiplets or new matter multiplets embedded in hypermultiplets 
(See \Sec{N2l}), $\ol{X}$ are the $N=2$ hyperpartners of chiral multiplet $X$, $S_X$ are the spectators needed to cancel the chiral anomaly, and $H_a$ stand for the matter multiplets in the $N=1$ sector; $N_F$ is 
the number of $L_i$, and also is that of $\ol{L}_i$ or $S_j$, i.e. $i$ runs from 1 to $N_F$; $N_f$ is the number of 
$H_a$; $M_W$ and $M$ are the Majorana SUSY mass for $W$ and $\f_Y$, respectively; $M_i$ is the Dirac SUSY 
masses for $\ol{L}_i$ and $S_i$;
$\tl{Y}_i$ and $\tl{w}_i$ are the $N=2$ Yukawa couplings (See \Sec{N2l}) in the unit of the corresponding gauge couplings.

$K_{N=2}$ and $W_{N=2}$ represent the K\"{a}hler and super- potentials of the same forms as those in $N=2$ SUSY QCD~\cite{Grimm:1977xp}, while $K_{N=1} \AND W_H$ are the ordinary $N=1$ K\"{a}hlar and super- potentials, 
respectively.  
Here, $W_H=W_H(H_a,S_j; L_i, \ol{L_i}, W, \f_Y)\simeq W_H(H_a,S_j)$ is a function of $H_a$ and $S_j$, and is assumed to weakly depend on $L_i$, $\ol{L}_i$, $W$, and $\f_Y$. 
Thus, we will neglect the dependence on $L_i, \ol{L}_i$, $W$ and $\f_Y$ of $W_H$.
In particular, we will neglect the Yukawa interactions for $L_i$, which can stand for the SM Yukawa couplings and breaks the $N=2$ SUSY explicitly.  The definitions are summarized in Table.\ref{tab:1}

\subsection{$N=2$ SUSY limit and Two Sectors}

\lac{N2l}
With the decoupling of the multiplets, $H_a$ and $S_j$,
\begin{equation}
\laq{N2l}
\w_i \-> 1, \tl{Y}_i \-> Y_i,
\end{equation}
is the $N=2$ SUSY limit~\cite{Grimm:1977xp}. 
To see the property at this limit, let us focus on some of the Yukawa couplings obtained from the $W_{N=2}$ and $K_{N=2}$:
\begin{equation}
\laq{su2R}
\mathcal{L} \supset \sum_{i=1}^{N_F}{\left( -i \sqrt{2}  \tl{L}_i^\* (g_c\l_c+Y_ig\l) \cdot \p_{Li} -\sqrt{2}  \tl{\ol{L}}_i ( \w_i g_c\psi_W +\tl{Y}_i g \psi)\cdot \psi_{Li}\right)}.
\end{equation}
Here, $\psi_W~ (\psi)$ and $\l_c~(\l)$ are the gaugini, the fermionic component of $W$ ($\f_Y$), and the gaugino in the adjoint representation of the $\SU(N_c)$ ($U(1)$) gauge groups;
$\tl{X} (\p_{X})$ are the scalar (fermion) components of $X$.
An $\SU(2)_R$ symmetry manifests itself in \Eq{su2R} with \Eq{N2l}, under which $\{ i\l_c, \psi_{W} \}$, $\{ i\l, \psi_{\f_Y} \}$ and $\{ \tl{L}_i,\tl{\ol{L}}_i^\*\}$ are doublets while the other fields participant in this limit are singlets. 
In the light of $\SU(2)_R$, we have two kinds of enlarged multiplets in $N=2$ SUSY, e.g. $N=2$ vector multiplets $\{V, \f_Y\}$ and $\{V_c, W\}$, and hypermultiplets $\{L_i, \ol{L}_i\}$. 

Since the first and second terms of \Eq{su2R} arise from the K\"{a}hler and super- potentials, respectively, the $\SU(2)_R$ transformation mixes the terms in these two potentials. 
Moreover, the K\"{a}hler potential and the SUSY gauge kinetic term are mixed due to the rotation of the components of $\{V, \f_Y\}$ or $\{V_c, W\}$. This fact will be essential to derive the $N=2$ non-renormalization theorem in \Sec{nonren}.

The particles present in the $N=2$ SUSY limit compose the $N=2$ sector,
namely $V_c, V, W, \f_Y$, $L_i$, and $\ol{L}_i$ are the components. On the other hand, the multiplets decoupling at this limit compose the $N=1$ sector, where $S_i$ and $H_a$ are the components.
Notice that the gauge couplings of $S_i$ and $H_a$, and Yukawa couplings in $W_H$ are hard breakings of the $\SU(2)_R$ symmetry.

We will use the definition made here to explain the phenomena in the $N=2$ sector, even with an explicit breaking of $N=2$ SUSY. The definition made here are summarized in Table.\ref{tab:1}.

\begin{table}[!t]
\begin{center}
\begin{tabular}{|c|c|c|c|c|c|c|c|c|c|c|c|}\hline
Sector&\multicolumn{4}{c|}{$N=2$ sector}&\multicolumn{2}{c|}{$N=1$ sector} \\\hline
Chiral multiplets&$W$&$\f_Y$&$L_i$&$\ol{L}_i$&$S_i$&$H_a$\\\hline
$N=2$ partner&$V_c$&$V$&$\ol{L}_i$&$L_i$&/&/\\\hline
$N=2$ multiplet&vector&vector&hyper-&hyper-&/&/\\\hline
SU($N_c$) representation&adjoint&1&$r_i$&$\ol{r_i}$&$r_i$&$r^H_{a}$\\\hline
U(1) charge&0&0&$Y_i$&$-Y_i$&$Y_i$&$Y^H_a$\\\hline
$i_{\rm max}/ a_{\rm max}$&/&/&$N_F$&$N_F$&$N_F$&$N_f$ \\\hline
\end{tabular}
\end{center}
\caption{Particle contents of a partially $N=2$ SUSY model \Eq{toy}. The representation $r_i$ is allowed to be unity. ``Vector" and ``hyper-" denote the corresponding particles are embedded in the vector and hyper- multiplets, respectively. $i_{\rm max} \AND a_{\rm max}$ denotes the maximum value of $i \AND a$, respectively. } \label{tab:1}
\end{table}

\subsection{Radiative Corrections}
\lac{RC}
The 1-loop RG equations for the dimensionless couplings in the $N=2$ sector are given as follows~\cite{Martin:1993zk, 2-loopRGE2}.
\begin{equation}
\laq{betag}
\dt{g_c}\equiv \beta_c={1 \over 16\pi^2} g_c^3 (F_2+F_1-2N_c), \ \dt{g}\equiv \beta={1 \over 16\pi^2} g^3 (f_2+f_1),
\end{equation} \begin{equation}
\laq{betaw}
\dt ( {g_c{w}_i})={g_c \w_i}(\gamma_W+\gamma_{Li}+\gamma_{\ol{L}i}), 
\end{equation} \begin{equation}
\laq{betaY}
\dt {(g \tl{Y}_i)}=g\tl{Y}_i(\gamma_{\f_Y}+\gamma_{Li}+\gamma_{\ol{L}i}),
\end{equation}
where the anomalous dimensions for $W, \f_Y, \AND   L_i,\ol{L}_i $ are given as
\begin{equation}
\gamma_{W}={1\over 16\pi^2}\left(\sum_i^{N_F}{2T({r_i}){\w}_i^2} -2N_c\right)g_c^2, \end{equation} \begin{equation} \gamma_{\f_Y}={1\over 16\pi^2}\sum_i^{N_F}{2d(r_i)\tl{Y}_i^2 g^2 },
\end{equation} \begin{equation}
\laq{gammai}
\AND \gamma_{Li}=\gamma_{\ol{Li}}={1\over 16\pi^2}\left(2({\w}_i^2 -1)C(r_i)g_c^2+2(\tl{Y}^2_i-Y_i^2)g^2\right),
\end{equation}
respectively. 
Here $t=\log{({\mu_{RG} \over \GEV})}$; 
$T(r)$, $C(r)$ and $d(r)$ denote the Dynkin index, the quadratic Casimir invariant and the dimension of the representation $r$, respectively; $\b_c$ and $\b$ are the 1-loop $\b-$functions for $\SU(N_c)$ and $\U(1)$, respectively; $F_2~(F_1)$ and $f_2~(f_1)$ are the sums of Dynkin indices of SU($N_c$) and U(1) in the $N=2~(N=1)$ sector,  respectively:
\begin{equation}
 \laq{def1}
F_2\equiv \sum_i^{N_F}{2T(r_i)}, \ F_1\equiv \sum_i^{N_F}{T(r_i)}+\sum_i^{N_f}{T(r^H_a)},
\end{equation} \begin{equation}
\laq{def2}
\AND f_2\equiv \sum_i^{N_F} 2d(r_i)Y_i^2, \ f_1\equiv \sum_i^{N_F}{d(r_i)Y_i^2}+\sum_i^{N_f}{d({r^H_a})(Y^{H}_i)^2}.
\end{equation}

At the $N=2$ SUSY limit, where $f_1=0,~ F_1=0$ with \Eq{N2l},
we evaluate $\dt{\w_i}=\dt{\tl{Y}_j}=\g_{Li,\ol{L}i}=0$ so that \Eq{N2l} and the vanishing of \Eq{gammai} are satisfied perturbatively at any scale.

\subsection{A Fixed Point of the $N=2$ Yukawa Couplings}

 \lac{fix} 
Let us investigate on the dimensionless couplings at the low energy analytically.

As shown in \Sec{RC}, at the limit of $N=2$ SUSY, \Eqs{N2l} are satisfied at any scale.  
This fact implies that the limit \eq{N2l} represents a fixed point in the parameter space characterized by $\{\w_i, \tl{Y}_j\}$. 
We will show that in the presence of the degrees of $H_a$ and $S_i$, the IR fixed point still exists and moves to a different position, $\{\ol{\w}_i, \ol{\tl{Y}}_j\}$.

First, we divide $N_F$ into $N_{F}^s$ and $N_{F}-N_{F}^s$, where $N_{F}^s$ is the total number of SU($N_c$) singlets in $L_i$ for convenience.
Without loss of generality, we can rearrange the indices of the $N=2$ sector, such that the superfields labeled by $i=1 \sim N_{F}^s$ are SU($N_c$) singlets, while those of $i=N_{F}^s+1 \sim N_{F}$ 
 are not. 
We also divide $f_2$ into
 \begin{equation}
 \laq{def3}
 f_2^s \equiv \sum_i^{N_F^s}{2Y_i^2}, \ f_2^{ns} \equiv \sum_{i=N_F^s+1}^{N_F}{2d(r_i)Y_i^2}.
 \end{equation}
 In the calculation we assume,
 \begin{equation} 
 \laq{univcond}
 C({r_i}) g_c^2 \sim C({r_i}) g_c^2 \w_i^2 \gg Y_i^2 g^2 \sim \tl{Y}_i^2 g^2.
 \end{equation} 
This condition stands for $(4/3)g_3^2 \OR (3/4)g_2^2 \gg Y^2_i g_Y^2$, where $g_Y, g_2,$ and $ g_3$ are the SM gauge couplings of $\U(1)_Y, \SU(2)_L$, and $\SU(3)$, respectively.

A solution of the vanishing condition for $\dt{\w_i}, \dt{(g\tl{Y}_i)}$ is,
\begin{equation}
\laq{fixw}
\ol{\w}_i^2-1\simeq {d(r_i) \over 2 T(r_i)}{F_1 \over \sum_{i=N_F^s+1}^{N_F}{d(r_i)}+2(N_c^2-1)}+\O({g^2 Y_i^2  \over g_c^2 C(r_i)}) ~~(i>N_F^s),
\end{equation}
\eqn{\laq{fixYns} g^2\ol{\tl{Y}}^2_i \simeq 0 ~~(i>N_F^s),}
The second equation is obtained by neglecting terms of $\O(g^2)$ while considering $(f_1+f_2) g^2$. This approximation stands for $(f_1+f_2) g_Y^2 > g_2^2 , g_3^2$ due to the large coefficient $f_1+f_2 \geq 11$ in the realistic case.

Hence we find a general relation,
\eqn{\ol{\w}_i^2-1>0.}

Now let us check whether $\{\ol{\w}_i, \ol{\tl{Y}}_i\} (i\geq N_F^s)$ with approximation \Eq{univcond} represents an IR fixed point. Rewriting \Eqs{betaw} and \eq{betaY} 
in terms of $\d \w_i^2 \equiv \w_i^2-\ol{\w}_i^2$ and $\d \tl{Y}_i^2\equiv \tl{Y}_i^2-\ol{\tl{Y}}_i^2$, we obtain the RG equations for the differences
\begin{equation}
\dt\d \w_i^2 \simeq \sum_{j=N_F^s+1}^{N_F} \d {\w}^2_j {g_c^2 \over 8\pi^2}A_{ij}, \non \end{equation} \begin{equation} \laq{betad} \dt (g^2 \d{\tl{Y}_i }^2) \simeq \sum_{j=N_F^s+1}^{N_F} g^2\d {\tl{Y}_j^2} {g_c^2 \over 8\pi^2}B_{ij} ~~(i>N_F^s)
\end{equation}
at the leading order. Here $A_{ij}$ and $B_{ij}$ are positive-definite matrices:
\begin{equation}
\laq{Aij}
A_{ij}=2\ol{\w}_i^2\left(T(r_j)+2C(r_j)\d_{ij}\right), \ B_{ij}=4(\ol{\w}_i^2-1)C(r_i)\d_{ij}.
\end{equation}
Employing the analytic solutions of \Eqs{betag}, we can solve \Eqs{betad} and obtain, 
\begin{align}
\laq{fixc1}
\d\w^2_i(\m_{RG}) &\simeq \sum_{j=N_F^s+1}^{N_F}{\left[\left({{1\over \a_{\rm p}} \over {4\pi \over g_c^2(\m_{RG})}}\right)^{{A \over F_1+F_2-2N_c}}\right]_{ij} \d \w^2_j(M_{\rm p})},\\
\laq{fixc12}
\d (g^2\tl{Y} ^2_i)|_{\m_{RG}} &\simeq  \sum_{j=N_F^s+1}^{N_F}{\left[\left({{{1\over \a_{\rm p}}} \over {4\pi\over g^2_c(\m_{RG})}}\right)^{{B \over F_1+F_2-2N_c}}\right]_{ij} \d(g^2\tl{Y}_j^2)|_{M_{\rm p}}}~~~ (i>N_F^s),
\end{align}
where $X|_\mu$ denotes the variable $X$ at the renormalization scale $\m$ and $\a_{\rm p}\equiv g_c(M_{\rm p})^2/4\pi$.
Thus, for $F_1+F_2-2N_c \neq 0$ \Eqs{fixw} and \eq{fixYns} represent an IR fixed-point. 
 
Following a same procedure, the $N=2$ Yukawa couplings for the $\SU(N_c)$ singlets have a fixed point, 
\begin{equation}
\laq{fixY}
\ol{\tl{Y}_i}^2-Y_i^2 \simeq {1 \over 2}\left( {f_1+f_2^{ns} \over N_F^s+2}\right) ~~(i\leq N_F^s),
\end{equation}
where we have set $Y_i =\ol{Y}_i , w_i=\ol{w}_i$ for $i>N_F^s$.
Hence,
\begin{equation}
\ol{\tl{Y}_i}^2-Y_i^2>0.
\end{equation}

The difference, $\d \tl{Y}_i^2\equiv \tl{Y}_i^2-\ol{\tl{Y}}_i^2$, at the scale $\m_{RG}$ is given by
\begin{equation}
\laq{fixc2}
\d\tl{Y}^2_i (\m_{RG}) \simeq  \sum _j^{N_F^s}{\left[\left({{{1\over \tl{\a}_{\rm p}}} \over {4\pi \over g^2(\m)}}\right)^{{C \over f_1+f_2}}\right]_{ij} \d \tl{Y}^2_j (M_{\rm p})}  ~~(i\leq N_F^s),
\end{equation}
 where
\eqn{C_{ij}=\ol{\tl{Y}}_i^2(1+\d_{ij})}
is a positive-definite matrix, and $\tl{\a}_{\rm p}\equiv g(M_{\rm p})^2/4\pi$.

In summary, the position $\{\ol{\w}_i, \ol{\tl{Y}}_j\}$ given by \Eqs{fixw},\eq{fixYns}, and \eq{fixY} in the parameter space represents an IR fixed-point at 1-loop order. Namely, at low energy the $N=2$ Yukawa couplings approach to $\{\ol{\w}_i, \ol{\tl{Y}}_j\}$ from the value around $\{1, {Y}_j\}$ at the partially breaking scale.

Therefore, a partially $N=2$ SUSY model has generally a striking feature. That is the typical pattern of Yukawa couplings controlled by IR physics and matter contents, which is insensitive to the partially breaking mechanism or the threshold corrections at $M_p$. 
The IR fixed point is easily reached in the realistic case because the additional matter contents enhance the gauge couplings at $M_p$ through the RG running and thus \Eqs{fixc1}, \eq{fixc12}, and \eq{fixc2} are suppressed.

\section{$N=2$ to $N=0$ SUSY breaking and Anomaly Mediation}

\lac{2}
The $N=2$ SUSY breaking to $N=0$ is turned on in this section, and we will show that if $\SO(2)_R$ remains after the SUSY breaking, the sfermion masses are forbidden by the $N=2$ non-renormalization 
theorem. 
The anomaly mediation effect at the previously discussed fixed point will be investigated. In particular, the tachyonic slepton problem is resolved automatically near the fixed point. 
We will also discuss the condition that suppresses the RG running effect so that the spectrum in the $N=2$ sector is almost induced by anomaly mediation.

\subsection{$N=2$ Non-renormalization theorem and Splitting Mass Spectra}
\lac{nonren}

Before a general discussion, let us consider a concrete model for $N=2$ SUSY breaking to $N=0$: an $N=2$ gauge mediation model \cite{Giudice:1998bp,Shimizu:2015ara}.
Suppose that the SUSY breaking are mediated by $N=2$ messengers, $\{\f_m, \ol{\f}_m\}$, which are introduced as hypermultiplets charged under $\U(N_c)$. 
The $N=2$ messengers are characterized by the superpotential,
\begin{equation} 
\laq{GMSB}
W_{SB}=\sqrt{2} \ol{\f}_m(\tl{Y}_m g \f_Y + \w_m g_c W+Z){\f}_m,
\end{equation}
where $Z\equiv M+\theta^2 F_Z$ is a SUSY breaking field with $M \gg \sqrt{F_Z}$, $M$ ($F_Z$) represents the messenger scale (SUSY breaking $F-$term), and $\tl{Y}_m, \w_m$ are the $N=2$ Yukawa couplings for the messengers. $Z$ can be identified as the vacuum expectation value for the chiral component of an abelian $N=2$ vector multiplet \footnote{ For example, a simple model where the $N=2$ vector multiplet has an $N=2$ Fayet-Iliopoulos term, $W=\x Z$ can induce an $F-$term to $Z$ spontaneously while the messenger scale is the scalar component of $Z$. (See \cite{PN2UV1} for reference.) If the messengers carry the charge of this abelian gauge group, then \Eq{GMSB} is obtained.}.
The $N=2$ SUSY limit of the Yukawa couplings is given as
\begin{equation}
\laq{N2lm}
\tl{Y}_m\->Y_m, ~\w_m\->1,
\end{equation}
where $Y_m$ is the $\U(1)$ charge of $\f_m$.

The soft mass squares for $L_i$ at the $N=2$ SUSY limit, induced by radiative corrections from the messengers, 
are given by \cite{Giudice:1998bp}
\begin{equation}
m_i^2={{1\o2}\ab{{F_Z\over Z}}}^2{(\dt{\g}_{i}^--\dt{\g}_{i}^+)}.
\end{equation}
where $i$ denotes $L_i, H_a$; $+(-)$ denotes the value evaluated above (below) $M$. 
Since \Eqs{N2l} and \eq{N2lm} are satisfied, from \Eqs{betag}-\eq{gammai}, we find
\begin{equation} 
m_{Li}^2=m_{\ol{L}i}^2=0.
\end{equation}

In fact, these vanishing masses are the consequence of symmetry and the holomorphy. From the non-vanishing expectation value, $F_Z$, in \Eq{GMSB}, the potential acquires,
\begin{equation}
\non
\d V= F_Z \tl{\f}_m \tl{\ol{\f}}_m +h.c. 
\end{equation} \begin{equation}
\laq{vec}
=  (\tl{\f}^*_m, \tl{\ol{\f}}_m)^* \cdot (\Re[F_Z]\sigma_1-\Im[F_Z]\sigma_2) 
\cdot (\tl{\f}^*_m, \tl{\ol{\f}}_m)^T,
\end{equation}
where $\sigma_1 \AND \sigma_2$ are the Pauli matrices while $T$ denotes the transpose.
Since \Eq{vec} represents an isovector in $\SU(2)_R \sim \SO(3)_R$ space, non-vanishing $F_Z$ breaks SUSY but preserves a $\U(1)_R \sim \SO(2)_R$ symmetry, which is a subgroup of $\SU(2)_R$ with rotating axis $\{\Re{F_Z},-\Im{F_Z}, 0\}$ in the isovector space. Thus, this symmetry must mix $\tl{\f}^*_m \AND \tl{\ol{\f}}_m$, which are anti-chiral and chiral scalar fields, respectively. The effective theory at low energy has this $\SO(2)_R$ symmetry.

In fact, the soft breaking mass squares from the k\"{a}hler potential, like
\begin{equation}
\d K\sim {\ab{Z}^2 \over M^2}{\left( L_i^\dagger e^{2 g_cV_c+2Y_i g V}L^i+\ol{L}_ie^{-2g_cV_c-2Y_i g V}\ol{L}^{i \dagger}\right)},
\end{equation}
is forbidden by the $\SO(2)_R$ symmetry and the holomorphy. This is because with the $\SO(2)_R$ symmetry the potential should include the term \Eq{su2R} multiplied by $\ab{Z}^2 \over M^2$ while the second term ${\ab{Z}^2 \over M^2} \tl{\ol{L}}_i ( \w_i g_c\psi_W +\tl{Y}_i g \psi)\cdot \psi_{Li}$ is never generated from the superpotential due to holomorphy. 
Thus, $\SO(2)_R$ symmetry and holomorphy forbid the soft breaking mass squares for the hypermultiplets. This is nothing but the consequence of the $N=2$ non-renormalization theorem for the wave function renormalization and is independent of the mediation mechanism.

Now come back to the setup of \Sec{1}. 
Suppose that the soft SUSY breaking terms are generated preserving the $\SO(2)_R$ symmetry above the partially breaking scale. 
Since is sequestered from the partially breaking sector at $M_{\rm p}$\footnote{We are assuming that the $\SO(2)_R$ breaking Yukawa couplings for the fields in the $N=2$ sector are small enough to be neglected, e.g.  $L_i$ are likely to be the first two generation sfermions, which have small SM Yukawa couplings. }, the $N=2$ sector has an approximate $\SO(2)_R$ symmetry. 
Therefore, the soft mass squares from the SUSY breaking of the 
sfermions are suppressed in the $N=2$ sector at $M_p$.

As the consequence of the approximate $\SO(2)_R$ symmetry in the $N=2$ sector at $M_{\rm p}$, 
the following relations among the parameters are obtained,
\begin{equation}
\laq{N2pcon1}
\w_i(M_{\rm p}) \simeq 1 ,~ \tl{Y}_i(M_{\rm p}) \simeq Y_i,
\end{equation}  
and 
\begin{equation}
\laq{N2pcon2}
m_{L_i,\ol{L}_i}^2(M_{\rm p}) \simeq 0.
\end{equation}

On the other hand, throughout this paper, we do not specify the parameters in the $N=1$ sector.

\subsection{Anomaly Mediation and Tachyonic Slepton Problem in the $N=2$ sector}

\lac{soft}

Since the SM is chiral such an $\SO(2)_R$ symmetry\footnote{It is difficult to introduce the chiral partners of the SM fermions to have an exact $\SO(2)_R$ symmetry. 
Since a fermion of the $N=2$ partner is charged under the SM gauge group, it should be heavier than $\O(100)\GEV$ due to the LEP and LHC constraints \cite{PDG}. 
A Dirac mass term between fermions of the SM and the $N=2$ partner is forbidden, otherwise the SM fermion would become too massive. 
A mass term between two chiral fermions in the $N=2$ partners is likely to be forbidden because the mass in which case should be generated via the EW symmetry breakdown, and cannot be too large due to the constraint from the precision measurement of the S parameter~\cite{PDG,Peskin:1990zt}. } should be broken by radiative corrections, and the sfermion masses in the $N=2$ sector are generated. 
One of the radiative corrections, which must be considered from supergravity, is anomaly mediation~\cite{ANSBLS, ANSBM}.
The scalar and gaugino masses induced by the anomaly mediation effect are given as,
\eqn{\laq{ANSB}m_{{Li},\ol{L}i, W,\f_Y}^2={1\o2}{m_{3/2}^2}\dt{\g_{{Li},\ol{L}i, W,\f_Y}},}
\begin{equation}
\laq{ANSBG}
M_c=m_{3/2}{\b_c \over g_c}, ~M=m_{3/2}{\b \over g}.
\end{equation}
where $m_{3/2}$ is the gravitino mass. Interestingly, this relation is a renormalization invariant of $N=1$ SUSY, and hence the anomaly induced mass is UV-insensitive. 
Notice that the anomaly induced mass squares of the sleptons in the MSSM, which have asymptotically non-free gauge interactions with small Yukawa couplings, are negative, i.e. the tachyonic slepton problem~\cite{ANSBLS}.

Substituting \Eqs{betag} and \eq{gammai} with the $N=2$ Yukawa couplings at the fixed point into \Eq{ANSB}, the anomaly induced mass for the partially $N=2$ SUSY model is obtained as 
\begin{equation}
m_{{L_i,\ol{L}_i}}^2|_{\rm fp} \simeq {1\o16\pi^2}m_{3/2}^2{(\ol{\w}_i^2-1)}C(r_i)\dt{g}_c^2+{1\o16\pi^2}m_{3/2}^2{(\ol{\tl{Y}}_i^2-Y_i^2)}\dt{g}^2\non \end{equation} 
\begin{equation} 
 = {1\o16\pi^2}m_{3/2}^2 \left({(N_c^2-1)F_1 \over \sum_{i=N_F^s+1}^{N_F}{d(r_i)}+2(N_c^2-1)} g_c \beta_{c}-2Y_i^2 g\b  \right) \non \end{equation}\begin{equation}
 \simeq {1\o16\pi^2}m_{3/2}^2 \left(
{(N_c^2-1)F_1 \over \sum_{i=N_F^s+1}^{N_F}{d(r_i)}+2(N_c^2-1)} g_c \beta_{c}\right),
\laq{anmdsmuon}
\end{equation}
for non-singlets of $\SU(N_c)$, while 
\begin{equation}
\laq{anmds}
\left. m_{{L_i,\ol{L}_i}}^2\right|_{\rm fp}\simeq {1\o16\pi^2}m_{3/2}^2{(\ol{\tl{Y}_i}^2-Y_i^2)}\dt{g^2} 
= {1\o16\pi^2}
m_{3/2}^2{ \left( {f_1+f_2^{ns} \over N_F^s+2}\right)}g\b,
\end{equation}
for singlets of $\SU(N_c)$. 
These masses are functions of the $g, g_c, m_{3/2}$ and the model constants, independent of the UV physics.  

We can see that these mass squares are always positive for positive $\b, \b_c$. 
Therefore, the sleptons in the $N=2$ sector do not have the tachyonic slepton problem near the fixed point.

We note that instead of the tachyonic slepton problem, the negative anomaly induced mass squared for the scalar component of $W$ may be generated. However, this
is not so problematic as the case of the sleptons since we are allowed to have a tree-level SUSY mass as in \Eq{mass}\footnote{In fact, \Eq{mass} includes the Majorana mass term for the gaugini, which breaks the $\SO(2)_R$ symmetry. This term may arise from supergravity without spoiling the approximate $\SO(2)_R$ symmetry at the global limit. This is because a holomorphic k\"{a}her term, like $K= c\tr[WW]$, vanishes and does not break $\SO(2)_R$ in global SUSY,  while in supergravity a gaugini Majorana mass term, like $W= c m_{3/2} \tr[WW]$, is generated.}.

\subsection{Anomaly Induced $N=2$ Sector}

\lac{andom}

An interesting possibility is that the masses of sfermions in the $N=2$ sector are dominantly induced by anomaly mediation and the spectrum for them becomes almost UV-insensitive.  
Since the anomaly mediation effect for the sfermion mass is at 2-loop order, we would like to find the condition suppressing the 1-loop and 2-loop RGE effects~\cite{Martin:1993zk, 2-loopRGE2}. The 1-loop RG effect can be suppressed if the following is satisfied,
 \begin{equation}
\laq{N2pcon3}
m_{W}^2(M_{\rm p})\simeq m_{\f_Y}^2(M_{\rm p})\simeq0,
\end{equation} \begin{equation}
\laq{21}
A-{\rm terms}|_{M_{\rm p}}\simeq0,~ M_c(M_{\rm p})\simeq M(M_{\rm p})\simeq0, 
\end{equation}
\begin{equation}
\laq{S}
S \equiv \sum_{i}^{N_f}{d(r^H_a)m_{H_a}^2{Y^H_a}}+\sum_{i}^{N_F}{d(r_i)(m_{L_i}^2-m_{\ol{L}_i}^2+m_{S_i}^2)Y_i}\simeq 0,
\end{equation}
 where $m_{X}^2$ is the soft mass of scalar $\tl{X}$; $M_c$ and $M$ 
 are Majorana masses for the gaugino of $\SU(N_c)$ and $\U(1)$, respectively; $S$ is the $D-$term of the $\U(1)$ gauge interaction.

\Eqs{N2pcon3} and \eq{21} can be obtained by simply assuming that the SUSY breaking spurion field $Z=M+\h^2 F_Z$ is charged under some hidden symmetry.
Obviously, the gaugino mass terms, e.g. \begin{equation} 
{Z\over M_{\rm p}}\tr[\mathcal{W}_c\mathcal{W}_c],\end{equation}
 are forbidden. Furthermore, the soft scalar masses of $W$ and $\f_Y$ are also suppressed due to the approximate $\SO(2)_R$ symmetry. 
This is because the soft scalar mass square for a vector partner, restricted by the $\SO(2)_R$ symmetry (similarly to the discussion in \Sec{nonren}), can be only originated from a kinetic term, like
\begin{equation} 
\d K={Z\over M_{\rm p}}\tr[{W^\dagger e^{-2g_cV_c} W e^{2g_cV_c}}] ,
\end{equation}
which is forbidden by the hidden symmetry.  

A vanishing $D-$term of $\U(1)$ is quite general in several mediation mechanisms, e.g. gauge mediation model, mSUGRA, etc. We do not discuss this further.

To suppress the 2-loop RG effect characterized by $({1\over 16\pi^2})^2  m_{H_a, S_i}^2$, 
\begin{equation} 
m_{H_a,S_i}^2\ll  m_{3/2}^2,
\end{equation}
should be satisfied. If $m^2_{H_a,S_i}$ are generated via partially breaking of $N=2$ to $N=1$ at $M_p$, $m^2_{H_a,S_i}$ may be suppressed to the gravitino mass, $m^2_{H_a,S_i} \sim \e m_{3/2}^2$ by assuming the order parameter of the partially breaking to be $\e=({M_{\rm p} \over M_{pl}})$ or assuming the further sequestering between the visible and the $N=2$ to $N=0$ SUSY breaking sectors.

In summary, we have explained a possible setup that the sfermion and gaugino masses are dominantly induced by anomaly mediation in the $N=2$ sector. 
Since the $N=2$ Yukawa couplings are controlled by the fixed point at the low energy, light sfermions and gauginos in the $N=2$ sector have a typical spectrum that is almost UV-insensitive.

\lac{sol}

\section{A partially $N=2$ SSM}

\lac{app}

We will calculate the 2-loop RG running of the couplings and the corresponding anomaly induced masses in a partially $N=2$ SSM numerically to confirm the previous fixed point phenomena at higher loop level. 
In particular, we take the condition discussed in \Sec{andom}, and suppose that the soft masses in the $N=2$ sector are dominantly induced by anomaly mediation.

We consider an example model where the additional particle contents satisfy gauge 
coupling unification at 1-loop level and the perturbativity of dimensionless couplings. 
We will take this restriction, but we do not impose that the additional particles or the $N=2$ multiplets in the $N=2$ sector should be embedded into complete GUT multiplets. This is because as proposed by \cite{GUTbreak} as a solution of the doublet-triplet and proton decay problems, the GUT breaking may be due to an orbifold projection in extra dimension which leads to missing GUT partners.
We take this possibility because an extra dimension scenario is one of the leading candidates for the partially breaking. This is also the reason that we will set the partially breaking scale to be the GUT scale,
\begin{equation}
M_p=2\times 10^{16}\GEV.
\end{equation}

\subsection{An Example Model}

\lac{case1}

Now let us introduce a partially SSM model. 
The two sectors are composed by
\bed[$N=2$ sector] $\{e_1,\ol{e}_1\}, \{e_2,\ol{e}_2\}, \{L_2,\ol{L}_2\}, \{V_{2},W \}$, and $\{V_{Y},\f_Y\}$.
\item[$N=1$ sector] $S_{L2}, S_{e1}, S_{e2}, G$, and $Q_{1,2,3}, u_{1,2,3}, d_{1,2,3}, L_{1,3}, e_{3}, H_u, H_d, V_3.$
\eed
Here, $X_i$ are the $i$th generation chiral multiplets including the SM fermion $\p_{X_i}$; $V_Y, V_2$, and $V_3$ are the MSSM gauge multiplets of $\U(1)_Y, \SU(2)_L$, and $\SU(3)$, respectively; $H_u$ and $H_d$ are the MSSM Higgs multiplets; 
$\ol{X}$ denotes the hyperpartner of the chiral multiplet $X$; $S_X$ are the spectator; 
$\f_Y$ and $W$ are the $N=2$ vector partners of the gauge multiplets, $V_Y$ and $V_2$, respectively;
$G$, which is an octet of the $\SU(3)$ and singlet under $\SU(2)_L \times \U(1)_Y$, is introduced to satisfy the gauge coupling unification at $M_{\rm p}$. 
The $N=2$ sector fields are chosen to be the first two generation sfermions because they have small SM Yukawa couplings, which may be due to the $\SO(2)_R$ symmetry. $L_2, 
e_2$ are assigned to be within the $N=2$ sector because there is a hint from the discrepancy of the muon $g-2$~\cite{muong2exp,muong2SM,muong2SM2} suggesting that $\tl{L}
_2, \tl{e}_2$ are light, while the absence of the electron electric dipole moment implies that one of the selectrons, $\tl{L}_1$, may be heavy.
To summarize, other than the MSSM particle contents, there are chiral multiplets of $\ol{L}_2, \ol{e}_2, \ol{e}_1$ as the hyperpartners of some of the leptons, the vector partners of $\f_Y, W$ to 
guarantee the partial $N=2$ SUSY, the spectators of $S_{L2},S_{e1}, S_{e2}$ for cancelling the chiral anomaly, and an $\SU(3)$ octet of $G$ for the gauge coupling unification. The superpotential is given as follows.
\begin{equation}
\laq{case1}
W=W_{\rm N=2} +W_{\rm mass} +W_{\rm MSSM},
\end{equation} \begin{equation}
\laq{case1sp}
W_{\rm N=2}=\sqrt{2}\ol{L}_2(\w_L g_2 W+\tl{Y}_L g_Y\f_Y) L_2 +\sqrt{2}\sum_i^2{\ol{e_i}g_Y \tl{Y}_{e_i} \f_Y e_i}
\end{equation} \begin{equation}
\laq{case1mass}
W_{\rm mass}=-M_G \tr[G^2]-M_W \tr[W^2]-M \f_Y^2-M_{L} \ol{L}_2 S_{L_2}-\sum_{i}^{2}M_{e_i}\ol{e}_i S_{e_i}
\end{equation} \begin{equation}
W_{\rm MSSM}\simeq y_t H_uQ_3 u_3+y_b H_d Q_3 d_3+y_{\tau} H_dL_3 e_3+\m H_u H_d. 
\end{equation}
Here $y_t, y_b$, and $y_{\tau}$ are the ordinary MSSM Yukawa couplings for top, bottom, and tau, respectively; $\mu$ is the Higgs 
mixing parameter; the other parameters are defined in analogy to those in 
\Eq{toy}.
 We have neglected the Yukawa couplings including the first two generation 
 fermions.

Following \Sec{soft}, the $\SO(2)_R$ symmetry and a hidden charge for the SUSY breaking field imply the following boundary conditions for the parameters at $M_{\rm p}$, 
\begin{equation}
\laq{inc1}
~ \w_{L} = 1,~ \tl{Y}_{L} = 1/2,~ \tl{Y}_{e_1} = \tl{Y}_{e_2} = 1,
\end{equation} \begin{equation}
~ m_{{e}_1}^2={1\o2}{m_{3/2}^2}\dt{\g_{e1}}, ~m_{{e}_2}^2 ={1\o2}{m_{3/2}^2}\dt{\g_{e2}},~
m_{{L}_2}^2 ={1\o2}{m_{3/2}^2}\dt{\g_{L2}}
\end{equation} \begin{equation}
~m_{{W}}^2={1\o2}{m_{3/2}^2}\dt{\g_{W}}~ m_{{\f}_Y}^2={1\o2}{m_{3/2}^2}\dt{\g_{\f_Y}},
\end{equation} \begin{equation}
 M_1= m_{3/2}{\b_Y \over g_Y}, M_2=m_{3/2}{\b_2 \over g_2}.
\end{equation}
and the $A-$terms in the $N=2$ sector are also assumed to be induced by anomaly mediation. 
$\b_Y\AND \b_2$ are the $\b-$functions of $g_Y$ and $g_2$, respectively; $M_1$ and $M_2$ are the bino and wino masses.
We do not specify the parameters in the $N=1$ sector except for the assumption of the vanishing $D-$term as discussed in \Sec{soft}. 
We also do not specify the SUSY Dirac and Majorana masses in \Eq{case1mass}. 

We assume the unification condition at the scale $M_{\rm p}$ with \eqn{ {3\over 5}4\pi/g_Y^2=4\pi /g_2^2=4\pi/g_3^2=1/\a_{\rm p} \simeq 11.5.}

\subsection{ The Low Energy Mass Parameters and the Yukawa Couplings }

\begin{figure}
\begin{center}
   \includegraphics[width=125mm]{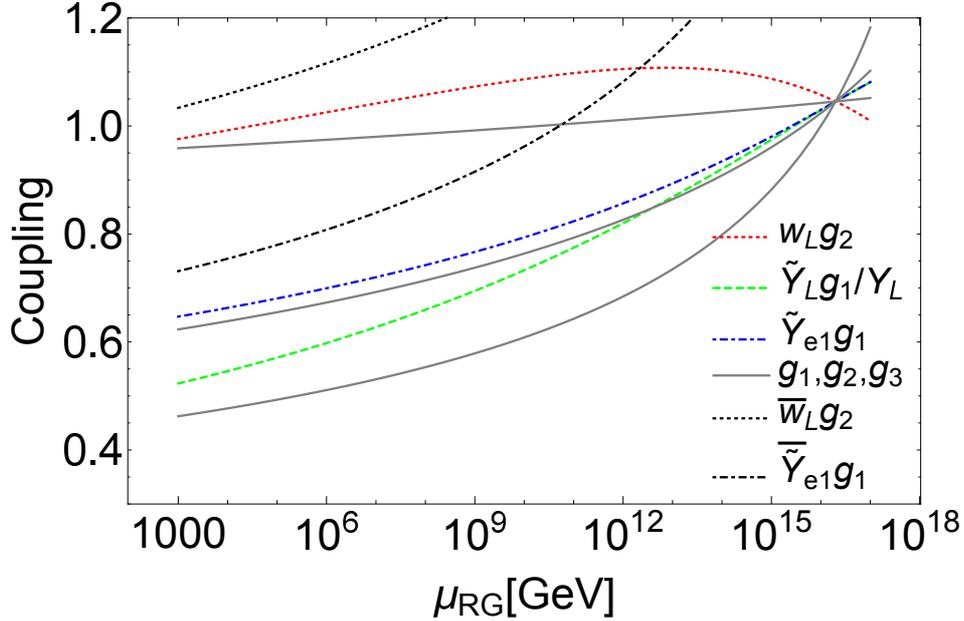}
  \end{center}
   \caption{ The 2-loop RG running of the gauge and Yukawa couplings. 
See the text for the details.}
 \label{fig:case1RG}
\end{figure}

We calculate the 2-loop RG equations with the boundary conditions of \Eq{inc1} where the 2-loop anomalous dimensions and the $\b-$functions are derived following~\cite{Martin:1993zk}.
The RG runnings of the relevant dimensionless couplings are illustrated in Fig.\ref{fig:case1RG}.
 The gray solid lines represent the scale 
dependence of the SM gauge couplings \{$\sqrt{5 \over 3}g_Y,g_2,g_3$\}, and the \{red dotted (black dotted), 
green dashed (zero-axis line), blue dot-dashed (black dot-dashed)\} lines represent that of \{$ g_2\w_{L}, 
2\sqrt{5 \over 3}g_Y \tl{Y}_{L}, g_Y\sqrt{5 \over 3}\tl{Y}_{e1, e2}$\} (at the fixed point), respectively. We can see that the $N=2$ Yukawa couplings approach to their fixed point values. 

\begin{figure}[t]
\begin{center}  
   \includegraphics[width=125mm]{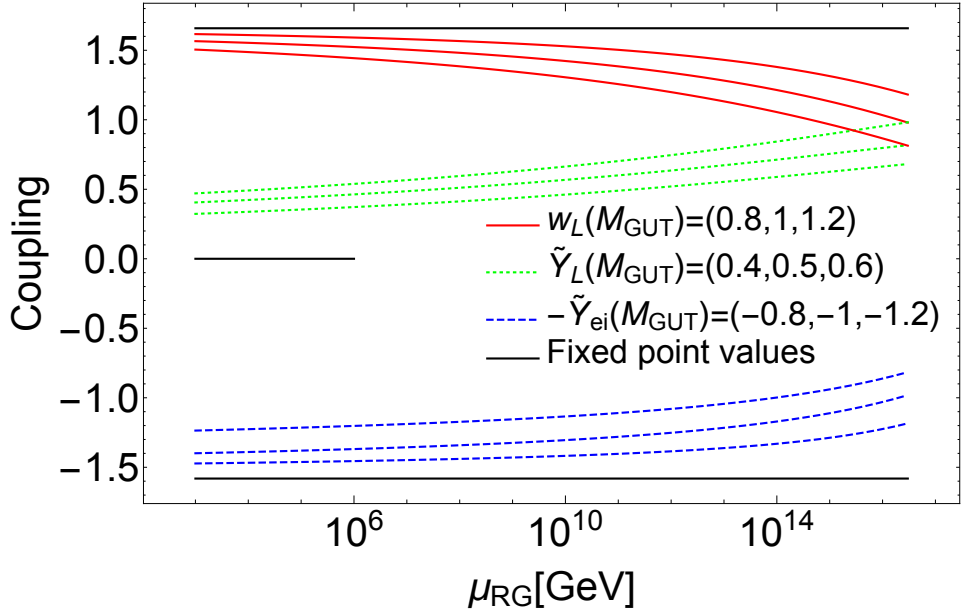}
      \end{center}
\caption{ The UV-insensitivity of the $N=2$ Yukawa couplings.  
 For illustrative purpose, we flip the sign for some parameters shown in the figure.}
 \label{fig:case1CV}
      \end{figure}

The convergence of the $N=2$ Yukawa couplings toward fixed point is shown in Fig. \ref{fig:case1CV}.
The red solid, green dotted and blue dashed lines are obtained with the boundary conditions:
$ 
\{\w_{L}, \tl{Y}_L, \tl{Y}_{e1}, \tl{Y}_{e2}\}= \{0.8 , {1\over 2}, 1,1 \} , \{ 1 , {1\over 2}, 1,1 \} ,\{1.2 , {1\over 2}, 1,1 \},$
$\{\w_{L}, \tl{Y}_L, \tl{Y}_{e1}, \tl{Y}_{e2}\}= \{1, 0.4, 1,1   \}, \{1, 0.5, 1, 1  \}, \{1, 0.6, 1,1  \},
$ $\AND  \{\w_{L}, \tl{Y}_L, \tl{Y}_{e1}, \tl{Y}_{e2}\}= \{1, {1 \over 2}, 1,0.8   \}, \{1, {1 \over 2}, 1, 1  \},\{1, {1 \over 2}, 1,1.2  \},$
respectively. The black solid line denotes the fixed point values of the $N=2$ Yukawa couplings evaluated by \Eqs{fixw}, \eq{fixYns}, and \eq{fixY}. Therefore, we find that the $N=2$ Yukawa couplings, especially $\w_L$, are almost UV-insensitive. Even if \Eqs{inc1} in the $N=2$ sector is more or less violated at $M_p$, the $N=2$ Yukawa couplings at the low energy are well approximated by these fixed point values, \Eqs{fixw}, \eq{fixYns}, and \eq{fixY}.

The numerically evaluated values (fixed point values at 1-loop order) of the couplings at the renormalization scale, $\mu_{RG}=$ 
10TeV, are
\begin{equation}
\w_{L}^2  \simeq 2~~({11\over 4}),~ g^2_Y\tl{Y}_{L}^2\simeq {0.05} ~~(0),~ \tl{Y}_{e_1}^2 \simeq \tl{Y}_{e_2}^2 \sim 2 ~~({5\over 2}), 
\laq{RGEc1}
\end{equation}
with 
\begin{equation}
g_Y^2\simeq 0.1,~ g_2^2\simeq 0.4,~ g_3^2\simeq 0.9.
\end{equation} 
Since the third generation and Higgs fields are in the $N=1$ sector, these values depend less on $y_t, y_b$ and $y_{\tau}$ and hence we do not specify $\tan\b$.

\begin{figure}   
\begin{center}
  \includegraphics[width=125mm]{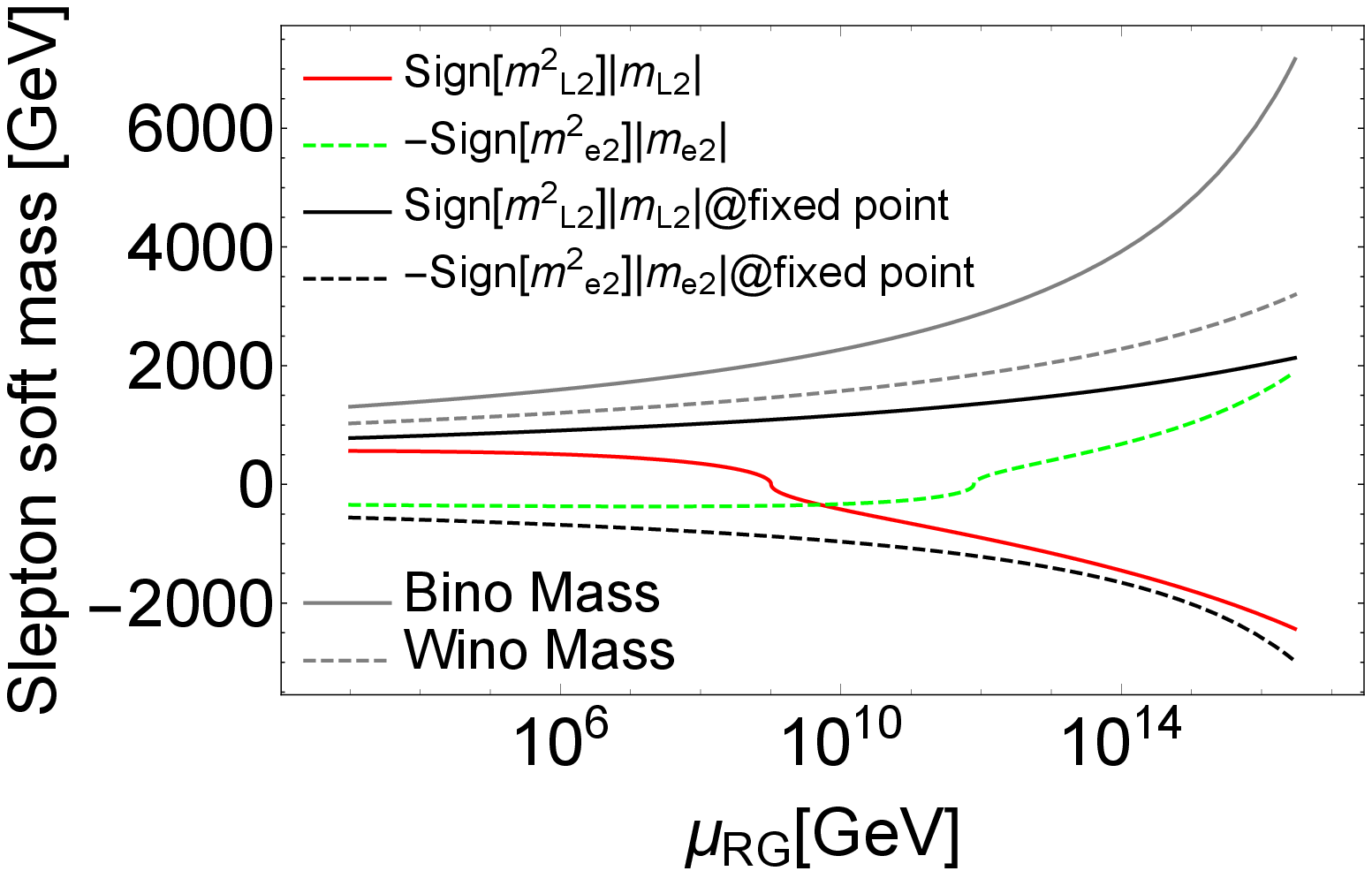}
        \end{center}
\caption{ The 2-loop RG running of the anomaly induced masses for the MSSM particles in the $N=2$ sector. 
 The sign of the vertical axis represents the sign of the mass squared. See the text.}
 \label{fig:case1RG2}
\end{figure}

In Fig. \ref{fig:case1RG2}, the scale dependence of some relevant anomaly induced masses is shown with $m_{3/2}=100\TEV$ at the 3-loop level (gaugino masses are evaluated at 2-loop level). The gray solid and dotted lines represent the 
scale dependence of bino and wino masses, respectively. The red solid (black solid) and green dashed (black dashed) lines represent ${\rm sign}({m_{{L}2}^2}) \ab{m_{{L}2}}$, \AND $-{\rm sign}({m_{{e}1}^2})\ab{m_{{e}_1} }\simeq -{\rm sign}({m_{{e}2}^2})\ab{m_{{e}_2}}$ (at the fixed point), respectively. Here, the sign of the vertical axis denotes the sign of the mass squared. We can see that below $\mu_{RG}=10^9 \GEV$ all the slepton mass squares become positive. 

\begin{figure}[t]
\begin{center}  
\includegraphics[width=125mm]{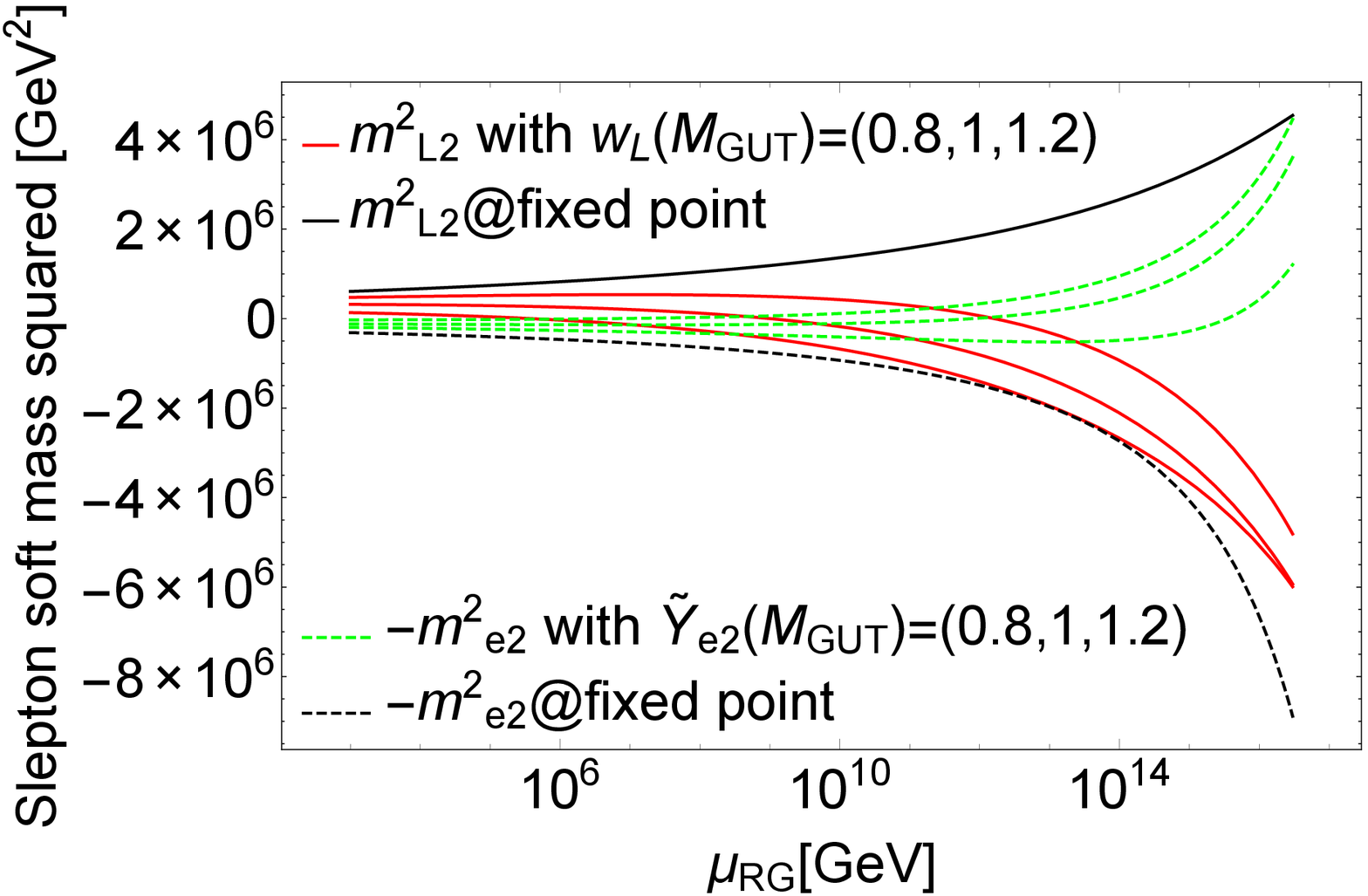}
 \end{center}
\caption{ 
The UV-insensitivity of the anomaly induced mass squares in the $N=2$ sector. 
 For illustrative purpose, we flip the sign for some parameters shown in the figure.}
 \label{fig:case1CV2}
 \end{figure}

 In Fig. \ref{fig:case1CV2}, we express the 
UV-insensitivity of these anomaly induced masses. The red solid and green dashed lines represent the runnings of the anomaly induced masses with the boundary conditions:
\begin{equation} \nonumber
\{\w_{L}, \tl{Y}_L, \tl{Y}_{e1}, \tl{Y}_{e2}\}= \{0.8 , {1\over 2}, 1,1 \} , \{ 1 , {1\over 2}, 1,1 \} ,\{1.2 , {1\over 2}, 1,1 \},
\end{equation}
\eqn{\nonumber
\AND  \{\w_{L}, \tl{Y}_L, \tl{Y}_{e1}, \tl{Y}_{e2}\}= \{1, {1 \over 2}, 1,0.8   \}, \{1, {1 \over 2}, 1, 1  \},\{1, {1 \over 2}, 1,1.2  \},}
respectively. The black solid (dashed) line represents the anomaly induced mass at the fixed point, \Eqs{anmdsmuon} and \eq{anmds}.

Therefore, we have shown that by using a concrete model the tachyonic slepton problem in the $N=2$ sector is solved in an UV-insensitive manner.

In particular, the numerical values of the relevant anomaly induced masses in the $N=2$ sector are evaluated at $\mu_{RG}=10~\TEV$,
\begin{equation}
\laq{spe}
m_{{e}1}=m_{{e}2}\simeq 0.004 m_{3/2}, ~m_{{L}_2}\simeq 0.006 m_{3/2},
\end{equation} \begin{equation}
\laq{spe2}
M_1\simeq 0.01 m_{3/2},~ M_2\simeq 0.01 m_{3/2}.
\end{equation}
Notice that the bino and wino are both heavier than the sleptons, and they cannot be the dark matter as in the ordinary MSSM. 
\subsection{Phenomenological Aspects}

\subsubsection*{\bf Tad-Pole Problem and Dark Matter Candidates}

Before discussing on the phenomenology, let us focus on a tad-pole problem due to the existence of the gauge singlet $\f_Y$~\cite{tadpoleprob}.
Since cubic radiative divergences are not forbidden in the softly broken SUSY theory, 
a fundamental singlet scalar has a large tad-pole term $ \sim m_{3/2}^2 M_{\rm p} \f_Y$, which would lead to a large vacuum expectation value of $\vev{\f_Y} \sim M_{\rm p}$.
This is quite problematic for our scenario because
the SM fermions embedded in the hypermultiplets become fairly massive through the $N=2$ Yukawa couplings.
A simple way to solve this problem is to impose an exact $Z_2$ symmetry\footnote{This symmetry could also be approximate but precise enough.}, under which $\f_Y$ is odd, and this term is forbidden. 
Then, we can find that $W$ should also be $Z_2$-odd. 
Thus, if a component of, $\f_Y \OR W$, is the lightest $Z_2$-odd particle, it can be the dark matter. 
Since this $Z_2$ is not necessarily an $R-$parity, the light sfermions can be even lighter than 
the dark matter, while with $R-$parity violation the lightest sfermion decays.
This possibility provides a significant feature of our scenario differing from the MSSM.
Therefore, we will consider that the $R-$parity is violated, while $Z_2$ symmetry is exact and a component of $\f_Y$ or $W$, the lightest $Z_2$-odd particle, is stable. 
From the superpotential \eq{case1}, the additional particles to the MSSM except for $G$ are all odd to guarantee the $Z_2$ symmetry. 

 The dark matter physics will be discussed later in detail.

\subsubsection*{\bf Muon $g-2$}

The sleptons in this scenario might be excluded up to $250\GEV$ ($350\GEV$) for right-handed ones (left-handed one)~\cite{Aad:2014vma}\footnote{The process we consider can be either slepton decaying to lepton and neutrino ($R-$parity violation) or slepton decaying to lepton and lightest singlet (it might be bini or wini). We have assumed the decoupling of the other particles absent in the process. For left-handed smuon, the bound should be over-estimated as we have only one light flavor. To the case of long-lived slepton the bounds see~\cite{ATLAS:2014fka}. }. Therefore, we may have a constraint,
\begin{equation}
m_{3/2}> 70\TEV.
\end{equation}

If the $\mu-$term, and the ratio of the vacuum expectation value of $H_u^0$ to $H_d^0$, i.e. $\tan\b$, are large enough,
the muon $g-2$ contribution can be evaluated as~\cite{muong2MSSM,Marchetti:2008hw} at the 1-loop level\footnote{ The additional particles contribute to the muon $g-2$ effective vertex at more than 2-loop level due to the $Z_2$ parity conservation. This is the reason we have used the formula for the MSSM.},
\begin{equation}
\laq{muong2}
\d\a_\mu \simeq \left(
\frac{1 }{1 + \Delta_\mu }
\right)
{g_Y^2 \over 16\pi^2}{ m_\mu^2  \mu \tan\beta \, M_1 \over m_{{L}_2}^2 m_{{e}_2}^2}
\,f_N\left( 
\frac{m_{{L}2}^2}{M_1^2},
\frac{m_{{e}2}^2}{M_1^2}
\right),
\end{equation} \begin{equation} 
 =25 \times 10^{-10}\left( {1.4 \over 1+\Delta_\m} \right) \left( { \mu\tan\b \over  300\TEV} \right) \left({80\TEV \over m_{3/2}}\right)^2,
\end{equation}
where, 
\begin{equation}
\Delta_\mu \simeq  \mu \tan\beta \frac{g_Y^2 M_1}{16\pi^2} I(M_1^2, m_{{ L}_2}^2, m_{{e}_2}^2)
\non
\end{equation} \begin{equation}
=0.4 \left( {\mu\tan\b \over 300\TEV}\right)\left( {80\TEV \over m_{3/2}}\right),
\end{equation}
and we have substituted the anomaly induced masses for the sfermions and the gauginos.
 Here, $I(x,y,z)$ and $f_N(x,y)$ are loop functions which can be found in the references. 
 If we quote~\cite{muong2exp, muong2SM}
\begin{eqnarray}
\d\a_{\rm exp} = (26.1 \pm 8.0) \times 10^{-10}.
\end{eqnarray}
 as a reference value of the experimental deviation from the SM prediction, 
 the muon $g-2$ anomaly can be explained within the error at the $1\s$ level ($2\s$ level)
 with $m_{3/2}\lesssim$100 TeV (120 TeV) for $\mu\tan\b =300\TEV$ for instance.

\subsubsection*{\bf Higgs Boson Mass}
On the other hand, the mass scale, $m_{N=1}$, of the $N=1$ sector can be much larger than that of the $N=2$ sector, depending on the detail of the partial SUSY breaking. 
Such a heavy scalar mass may raise the sfermion masses in the $N=2$ sector via 2-loop RG running,
the contribution of which can be approximated  by~\cite{Martin:1993zk,2-loopRGE2}
\begin{equation}
\non
\d m_{{L}_1,{e}_{1,2}}^2 \sim {(g_2^2  \OR g_Y^2)\over (16\pi^2)^2} \ln{\left({m_{N=1}\over M_{\rm p}}\right)} \left(g_3^2 \OR y_t^2\right) m_{N=1}^2 \end{equation} \begin{equation}
\sim \left(400 \GEV\right)^2  \left({m_{N=1} \over 40\TEV}\right)^2.
\end{equation}
Here we have used the fact that the $N=2$ sector fields are color singlets of $\SU(3)$, and hence the RG effect must contain EW gauge couplings or the $N=2$ Yukawa couplings which are of the same order.

Therefore, the parameter region, where the $N=2$ sector sfermion masses are dominantly induced by anomaly mediation, can be roughly characterized as 
\begin{equation}
\laq{upperbound}
m_{N=1} \lesssim 40 \TEV \left( {m_{3/2} \over 100\TEV}\right).
\end{equation}

Since having a small threshold correction due to the 
suppressed mixing term of the stops, the Higgs mass is mainly raised by the RGE effect through the top-loop which is cut-offed by the stop mass\cite{Okada:1990vk}, and thus a large stop mass scale is predicted. 
By using FeynHiggs~\cite{feynhiggs}, we found that $m_{\rm stop} \gtrsim 6 \TEV$ is obtained for $\tan\b \gg 10$  for the typical spectra where squarks, higgsino, MSSM Higgs are heavy while gauginos are light.\footnote{In the estimation, we have not included the effect of the additional particles which is at 2-loop order.} Thus the stops in the $N=1$ sector are allowed to give a large quantum correction to explain the correct Higgs boson mass \cite{Okada:1990vk}.
Unfortunately, a large stop mass implies we have several amounts of tuning to obtain the correct EW vacuum.

\subsubsection*{\bf Constraints from Particle Physics and Cosmology}

Paying the cost of the tuning, we can obtain several relaxations for the ordinary problems of the MSSM 
simply due to the heavy $N=1$ sfermions and the gravitino.
In the light of the heavy sfermions in the $N=1$ sector, SUSY contributions to the 
FCNC and CP-violating processes are suppressed\footnote{We have assumed the flavor symmetry conservation for the $N=2$ sector field for this example model.}. 
Furthermore, the heavy gravitino, $m_{3/2}\sim 100 \TEV$, decays much earlier than the BBN era and the cosmological gravitino problem is alleviated~\cite{Kawasaki:2008qe}. 
Also, the SUSY breaking field is not a singlet, and we do not have the cosmological moduli/Polonyi problem \cite{Polonyi}.

One may worry about the vacuum decay problem because we have large trilinear terms proportional to $\m\tan\b$, which implies the existing of charge breaking deeper minima than the EW vacuum of the potential. Since the fields in the $N=1$ sector are heavy and can be neglected in the discussion, the EW vacuum dominantly decays into the smuon number violating one. This was studied in \cite{Endo:2013lva}, from which we get $\m \tan \b<1140\TEV (1400\TEV)$ with $m_{3/2}=100\TEV (120\TEV)$ in our case. 
\\

Let me comment on a problem due to the additional color octet and its possible solutions.
The anomaly induced mass for gluino vanishes at 1-loop level, and is generated at 2-loop level as
\begin{equation}
M_3 \sim 0.002 m_{3/2},
\end{equation}
at $\m_{RG}=10\TEV$.
The gluino mass, if given by this formula, is too small to survive the 
experimental constraints with $m_{3/2} =\O(100)\TEV$ \cite{ATLAS:2016nij}.

There are two ways to tackle this problem. One is to introduce a Dirac gluino mass term, as $W={{D^a \over 
M_{\rm p}}\tr[G W^{(3)}_a]}$, where $D^a=D_Z \h^a$ and $W^{(3)}_a$ are a spurion SUSY breaking 
field\footnote{This is a natural introduction when the SUSY breaking $N=2$ multiplet is a vector multiplet, 
as $D^a$  can be recognized as field strength for the vector partner of the chiral SUSY breaking field, 
$Z$. } 
 and the field strength of the $\SU(3)$ gauge interaction, respectively. Notice that such a Dirac mass term 
 is not allowed for bino or wino due to the $Z_2$ parity, and our prediction would not change unless $D_Z$ 
 is extremely large. 
 The other way is to have a large $M_G$ with a supergravity induced ``$b-$term", $V= M_Gm_{3/2}\tr[GG]$. Then the decoupling of $G$
 induces a gauge mediation effect \Eq{GMSB} to raise the gluino mass to be the MSSM anomaly induced 
 one, $\gtrsim 2 \TEV$, with $m_{3/2}\gtrsim 100\TEV$. The $N=2$ sector spectrum does not change at the leading order.
In the Higgs mass calculation, we have taken the latter possibility.

\subsubsection*{\bf Dark Matter and Leptogenesis}

For simplicity, suppose that the lightest $Z_2$-odd particle is much lighter than the other additional particles to the MSSM so that we can discuss the dark matter physics in a generic manner. 
If the dark matter is wini (bini), i.e. the fermionic component of $\f_Y$ ($W$), it does not have any Yukawa interactions due to this assumption.
The physics of wini dark matter is similar to the pure wino case~\cite{Ibe:2011aa}, the difference of which will be discussed in \Sec{con}, while the bini is decoupled from the SM sector.

The interesting candidates of dark matter are the sbino and swino, i.e. the scalar components of the $\f_Y$ and $W$, respectively. For instance, let us consider the imaginary part of sbino, $\f\equiv {1\over \sqrt{2}}\Im{\tl{\f}}_Y$. 
It has a quartic potential only with the $N=2$ sector sfermions given by
\begin{equation}
\laq{quart}
V_\f \sim  \sum_{i=1}^{2}{\left({m_{{S}_{ei}}^2}\over m_{{S}_{ei}^2+M_{S_{ei}}^2}  \right)g_Y^2\ab{\tl{Y}_{e_i} \tl{e}_i \f}^2} +\left({m_{{S}_L}^2 \over m_{{S}_L}^2+M_{{S}_L}^2}  \right) g_Y^2\ab{\tl{Y}_{L} \tl{L}_2 \f}^2
\end{equation}
The ratio of the mass terms denotes the effect of the non-SUSY decoupling of the spectators (we have neglected the ``$b-$terms" for illustrative purpose), and at the SUSY limit, this vanishes.
However, this becomes an $\O(1)$ coefficient in general due to the SUSY breaking terms.

Now let us discuss the abundance of the dark matter.
Since at the fixed point $\tl{Y}_{L}=0$, the first term of \Eq{quart} denotes the dominant interaction for the sbino and thus controls the annihilation process represented by
\begin{equation}
\f+\f \-> \tl{e}_i + \tl{e}^\*_i.
\end{equation}
Since in the early universe the annihilation of the dark matter occurs only when $\tl{e}_i$ 
is lighter than the dark matter,
\begin{equation} m_{{e}_i}<m_{\f}\end{equation} 
is required. 
The thermal averaged total cross section is approximated by 
\begin{equation}
\vev{\s_{\f\f}\ab{v} }\sim {1 \over 8\pi} {\sum_i^{2}{\ab{g_Y^2 \tl{Y}^2_{e_i}}^2}\over m_\phi^2} \simeq (0.1/\TEV)^2 \left({700\GEV \over m_\f}\right)^2,
\end{equation}
where we have used the fixed point value of $\tl{Y}^2_{e_i}=5/2$ in the second approximation.
Thus the thermal abundance given is approximated by
\begin{equation}
\Omega_{\rm th} h^2 \simeq 0.1 \left({0.01 \TEV^{-2} \over \vev{\s_{\f\f}\ab{v} }}\right) \sim 0.1 \left({m_\f \over 700\GEV}\right)^2,
\end{equation}
compared with the observed dark matter abundance $\Omega_{\rm DM}h^2\simeq 0.12$ \cite{Ade:2015lrj}. Notice that the over-abundant problem, which should be cared for in the ordinary case of the bino-like neutralino dark matter, is avoided in the sbino case due to the light annihilation products of sleptons and the large quartic couplings.

As the ordinary SUSY, the heavy gravitino can decay into dark matter, which contributes to dark matter abundance non-thermally~\cite{Ibe:2011aa}.
The contribution in our scenario is 
\begin{equation}
\laq{nonth}
\Omega_{\rm nt}h^2\simeq {\rm 2Br}_{Z_2} \Omega_{3/2}h^2 {m_{\f}\over m_{3/2}} \end{equation} \begin{equation}
\sim 0.16 \left({2 n^{\rm odd}_{\chi}/12 \over n_V+n_{\chi}/12}\right) \left( {m_\phi \over 300\GEV}\right) \left({T_R \over 10^{10} \GEV}\right).
\end{equation}
Here $T_R$ is the reheating temperature; ${\rm Br}_{Z_2}$ is the branching ratio of the gravitino decay to the $Z_2$ odd particles; $\Omega_{3/2}$ is the energy density of the gravitino before its decay.
$n_V$ ($n_\chi, n_\chi^{\rm odd}$) is the effective number of the vector (chiral, $Z_2$-odd chiral) multiplet, and is 1+3+8, (49+12+8, 12).
There is a suppression factor of ${\rm Br}_{Z_2}$ because the gravitino is $Z_2$-even and it rarely eventually decays into the dark matter when the direct decay products are $Z_2$-even particles. 
Thus, the abundance should be multiplied by $2{\rm Br}_{Z_2}$.\footnote{The additional chiral multiplets do not change the thermally produced gravitino abundance at the leading order because the production is dominated by the dimension-5 gravitino-gauge interaction~\cite{Moroi:1993mb} which does not differ from the MSSM one in our scenario.}

In summary, we find that for 
\begin{equation} m_\phi\lesssim 700\GEV \end{equation} 
the correct dark matter abundance $\Omega_{DM}h^2 = \Omega_{\rm th}h^2+\Omega_{\rm nt}h^2$ can be obtained with a certain $T_R \gtrsim 2\times 10^{10}\GEV$.

In particular, thermal leptogenesis \cite{Fukugita:1986hr} requires $T_R\gtrsim10^{9.5}\GEV$ \cite{Ibe:2011aa}.
We conclude that our scenario is compatible with thermal leptogenesis.

\subsubsection*{\bf Predictions}

The direct detection constraints should not be stringent.
This is because the sbino couples to a nucleon with a spin-dependent suppressed interaction through a Z-boson boson coupling induced by a slepton loop. 
The direct and indirect detections in detail will be discussed elsewhere. \\

Since the annihilation of the sbino dark matter is viable only when an $N=2$ slepton is lighter than its mass,
the dark matter mass range turns to be a robust prediction of our scenario, that is
\begin{equation}
m_{{e}_1,{e}_2} \lesssim 700\GEV.
\end{equation}
Notice that the muon $g-2$ anomaly can be explained at the $1\s$ level with a certain $\m \tan\b$ in this mass range satisfying all the constraints discussed above.
If the $R-$parity violating decay of the light slepton occurs out of the detector, 
this mass range can be fully tested in the LHC with Drell-Yang production 
in a spectrum independent manner \cite{Heisig:2011dr}. If they decay within the detector they could be also tested as the $R-$parity violating scenario \cite{Aad:2014vma}.

From the relation of the anomaly mediation, we predict the upper bound of 
\begin{equation}
m_{{L}_2}\lesssim 1.1 \TEV \AND M_{1,2} \lesssim 1.8\TEV.
\end{equation}
The chargino with this mass range can also be produced in the LHC and would be followed by 
a typical decay to smuon and muon neutrino ( $\chi^-_1\-> \tl{\m}^{-}_2+\n_2$), or smuon neutrino and muon( $\chi^-_1\-> \tl{\n}_2+\m^-_2$). In this case, our scenario can be confirmed by measuring the typical mass pattern, especially for that for wino and light sleptons.

On the other hand, the $N=1$ sector sfermions as well as the $N=2$ partners, if are light enough, can be produced in high energy future colliders, such as FCC, SPPC, CLIC and a Muon Collider Higgs Factory \cite{CEPC}. 
In particular, if the mass of a fermionic hyperpartner of electron (muon) is within the reach in the electron (muon) collider, 
the hyperpartners are pair-produced via sbino propagating t-channel process, $e^{-}+e^{+}\-> \ol{e}^++\ol{e}^-(\m^{-}+\m^{+}\->  \ol{\m}^++\ol{\m}^-$).
In this case, the production rate is proportional to the fourth power of the $N=2$ Yukawa coupling. Thus, the typical Yukawa coupling controlled by the fixed-point can be obtained if the production rate is carefully measured, which could be a striking evidence of our scenario.

\section{Discussion and Conclusions}

\lac{con}

\lac{discuss}

Since essentially relying on the behavior around an IR fixed point, several discussions can apply to the models without partially $N=2$ SUSY but with adjoint chiral multiplets. On the other hand, we may consider the possibility that some of the quarks are within the $N=2$ sector, which can lead to light squarks and gluinos. 

I would like to mention that the difference of the pure wini dark matter and the pure wino one in the light of the leptogenesis. 
The thermal abundance estimation of the pure wini dark matter is quite similar to the pure wino case, the mass of which is bounded up to 2.7TeV from the correct abundance.
It is known that the ordinary pure wino dark matter should have $M_{2}\lesssim 1\TEV$ to be compatible with thermal leptogenesis \cite{Ibe:2011aa}. This is bounded from the non-thermal production of dark matter abundance. 
In our case, thanks to the suppression factor in \Eq{nonth}, the pure wini dark matter has a larger mass range compatible with thermal leptogenesis, which is $M_{W}\lesssim 2.7\TEV$ for the example model.\\

We have investigated on partially $N=2$ supersymmetric (SUSY) extensions of the standard model, which is composed of two 
sectors with almost $N=2$ SUSY and $N=1$ SUSY at the partially breaking scale, respectively.
Since the global $N=2$ SUSY is expected from the simple toroidal compactification of extra-dimensional SUSY theory, 
including the effective theory of superstring, the SM may be originated from $N=2$ SUSY. If the partially breaking of $N=2$ to $N=1$ takes place in a sequestered sector, the $N=2$ partners in the $N=2$ sector may be light enough to give interesting low energy phenomena. 
In particular, we have shown that the light $N=2$ partners in the $N=2$ sector has an almost UV-insensitive significant Yukawa interactions due to an IR fixed point.

From these interactions, the typical anomaly induced masses for the sfermions and the gauginos in the $N=2$ sector are almost UV-insensitive. 
In fact, we have clarified that the anomaly mediation goes well with the partially $N=2$ SSMs in two aspects:
(a) partially $N=2$ SUSY can provide a good condition for the anomaly mediation to be effective in the $N=2$ sector due to the $N=2$ non-renormalization theorem, and (b) the tachyonic slepton problem is automatically solved due to the large $N=2$ Yukawa couplings around the IR fixed point.

In a concrete model of a partially $N=2$ SSM with gauge coupling unification, we have shown that the muon $g-2$ anomaly can be explained within its 1$\s$ level error by the light smuons and gauginos. The masses are anomaly-induced and are 1loop suppressed to the gravitino mass of $\O(100)\TEV$. 
We have discussed the phenomenological and cosmological aspects.  
In particular, we have considered the dark matter candidate as sbino, the scalar component of the singlet $N=2$ vector multiplet. 
To explain the dark matter abundance, the sbino and right-handed smuon are required to be lighter than 700GeV. 
Then from the anomaly mediation relation we predict that all the sparticles of the MSSM in the $N=2$ sector have masses below $\sim 2\TEV$ with a pattern.
These are our robust predictions and can be tested and could be confirmed in the LHC or in the future colliders. 
If the $N=2$ partners are also reachable in the future colliders, the predicted $N=2$ Yukawa couplings might be measured as a striking evidence of the scenario.

\section*{Acknowledgments}
I thank Tetsutaro Higaki and Kazuhiro Endo for carefully reading this manuscript and for helpful suggestions.
I am also grateful to Tetsutaro Higaki and Norimi Yokozaki for useful discussions. 

\providecommand{\href}[2]{#2}\begingroup\raggedright\endgroup

\clearpage

\end{document}